\documentclass[12pt]{article}
\usepackage{amsmath, amsfonts, amssymb}
\usepackage[cp1251]{inputenc}
\usepackage[english,russian]{babel}
\usepackage{graphicx}
\usepackage{indentfirst}
\usepackage{threeparttable}
\usepackage{url}

\begin{document}
\title{\LARGE \bf VARIABLE STARS: A NET OF COMPLEMENTARY METHODS FOR TIME
SERIES ANALYSIS. APPLICATION TO RY UMA}
\author{\bf I.~L.~Andronov$^1$\thanks{E-mail: tt\_ ari@ukr.net} , L.~L.~Chinarova$^{2,1}$\thanks{E-mail: llchinarova@gmail.com}}
\date{\it  \small  $^1$ Odessa National Maritime University, \\
Mechnikova st. 24, Odessa, 65029 Ukraine\\
 $^2$ Odessa I. I. Mechnikov National University, \\
Dvorianskaia st. 2, Odessa, 65000 Ukraine
}
\maketitle

\renewcommand{\abstractname}{}
\begin {abstract}
\bf Abstract \rm  -- 
The expert system for time series analysis of irregularly spaced signals is reviewed. It consists of a number of complementary algorithms and programs, which may be effective for different types of variability. Obviously, for a pure sine signal, all the methods should produce the same results. However, for irregularly spaced signals with a complicated structure, e.g. a sum of different components, different methods may produce significantly different results. 

The basic approach is based on classical method of the least squares (1994OAP.....7...49A). However, contrary to common ''step-by-step'' methods of removal important components (e.g. mean, trend (''detrending''), sine wave (`'prewhitening''), where covariations between different components are ignored, i.e. erroneously assumed to be zero, we use complete mathematical models. 

Some of the methods are illustrated in the observations of the semi-regular pulsating variable RY UMa. The star shows a drastic cyclic change of semi-amplitude of pulsations between 0.01 to 0.37mag, which is interpreted as a bias between the waves with close periods and a beat period of 4000d (11yr). The  dominating period has changed from 307.35(8)d before 1993 to 285.26(6)d after 1993. The initial epoch of the maximum brightness for the recent interval is 2454008.8(5). It is suggested that the apparent period switch is due to variability of amplitudes of these two waves and an occasional swap of the dominating wave.

\end{abstract}

\section{Introduction}

Variable stars represent very different types of signal shapes. The official classification is published in the `'General Catalogue of Variable Stars'' (GCVS)  (Samus et al., 2017). In the current (September, 2019) version of the electronic catalogue of GCVS, there are 78 073 variable stars, which are classified into 569 combinations of 44 types. So many of these stars show multi-component variability.

The mathematical modelling of signals may be split into the `'physical'' and `'phenomenological'' methods. The first group tries to determine the physical parameters by fitting the theoretical curve to the observations. However, often the number of physical parameters is much larger than may be determined from the current observations. For example, by determining visual brightness (one parameter), one may not determine the absolute brightness and the distance (two parameters). Other examples may be the size of the spot in the atmosphere and its relative brightness,  etc.

So there may be `'observational facts'', or `'phenomenological parameters'', which may be then used in further models with an additional information.

In this short review paper, we list our main methods and show related references to the original papers. The illustration is presented for the semi-regular variable RY UMa.

\section{Basic Methods}

\subsection{Test Function}
The common method of the parameter determination is to minimize (sometimes, maximize) the test function $\Phi(x_k;t_k;C_\alpha),$ which is dependent on the observations $x_k,$ $k=1..n$ obtained at times $t_k,$  and on a set of parameters $C_\alpha,$ $\alpha=1..m.$ From the statistical point of view, the parameters should maximize the likelihood function (Anderson, 2003). Under a common assumption that the statistical errors $\sigma_k$ of the observations $x_k$ are random numbers with a zero mathematical expectation (i.e. no systematic shifts) and normally distributed, the test function is typically defined as
\begin{equation}
\Phi_m(x_k;t_k;C_\alpha)=\sum_{k=1}^n w_k\cdot(x_k-x_C(t_k;C_\alpha))^2,
\end{equation}
where $x_C(t_k;C_\alpha)$ is the `'computed'' value for a given argument $t_k$ and coefficients $C_\alpha.$ 
The `'weights'' $w_k$ are to be defined as $w_k=\sigma_0^2/\sigma_k^2,$ where the `'unit weight'' error $\sigma_0$ may be, in principle, be any constant positive value. Often the programs (e.g. electronic tables) neglect the possible difference in weights, what is equal to set all of them $w_k=1.$

In the `'linear least squares'' method, the approximation 
\begin{equation}
x_C(t;C_\alpha)=\sum_{\alpha=1}^m C_\alpha\cdot f_\alpha(t),
\label{xc}
\end{equation}
where $f_\alpha(t)$ are called the `'basic functions''. For the `'non-linear least squares'', at least some of the basic functions are dependent on the coefficients. In this case, the test function is computed at a grid of values of these `'non-linear coefficients'', the position of the minimum is used as an initial `'vector'' (set of values) and then is corrected to more accurate values using the `'differential corrections'' (see Andronov, 1994a, 2003, 2020,  Andronov and Marsakova, 2006 for more details). In this case, the basic functions may be extended to a definition
\begin{equation}
f_\alpha(t)=\frac{\partial x_C(t;C_\alpha)}{\partial C_\alpha},
\end{equation}
The variance of the approximation is
\begin{equation}
\sigma^2[x_C(t;C_\alpha)]=\sigma^2_M\cdot \sum_{\alpha\beta=1}^m A_{\alpha\beta}^{-1}f_\alpha(t)f_\beta(t),
\end{equation}
where $\sigma^2_M=\Phi_M/(n-M),$ and $M$ is a complete number of the parameters, including $m$ `'linear'' parameters. The matrix of normal equations
\begin{equation}
A_{\alpha\beta}=\sum_{k=1}^n w_k\cdot f_\alpha(t_k)\cdot f_\beta(t_k).
\end{equation}

\subsection{Multi-Component Signals}

Complicated models may be subdivided into `'linear'' (just a sum of larger summands in Eq. (2)) or `'non-linear'' ones. To determine the parameters, it is natural, in both cases, to vary a complete set of parameters. However, often more simple models are used, which are applied consequently.'

Contrary to common ''step-by-step'' methods of removal important components (e.g. mean, trend (''detrending''), sine wave (`'prewhitening''), where covariations between different components are ignored, i.e. erroneusly assumed to be zero, we use complete mathematical models. 

Generally, the matrix $A_{\alpha\beta}$ is not diagonal (i.e. the basic finctions are not orthogonal), so not diagonal is the inverse matrix $A_{\alpha\beta}^{-1}.$
This is often neglected, and the solutions and error estimates may significantly differ from the statistically optimal ones.

The oversimplification of the expressions was called the ''matrix-phobia'' by Prof. Z. MikulГЎЕЎek (2007). It may change the estimates of the parameters by a few dozen percent, and, in worst cases, by a factor of few times or even dozens times.

\subsection{Periodogram analysis}

For the periodogram analysis, we use a trigonometrical polynomial model of order $s$ (up to $(s-1)$-th harmonic), which is added to an algebraic polynomial of order $q$:
\begin{equation}
x_C(t;C_\alpha)=\sum_{\alpha=1}^{q+1} C_\alpha\cdot t^{\alpha-1}+\sum_{j=1}^s (C_{2j+q}\cdot\cos(j\omega t)+C_{2j+q+1}\cdot\sin(j\omega t) ,
\label{tp}
\end{equation}
where $\omega=2\pi f,$ $f=1/P$ is a trial frequency corresponding to a trial period $P.$ As the test function for the periodogram, we use the ratio
\begin{equation}
S(f)=1-\frac{\Phi_{q+1+2s}}{\Phi_{q+1}}
\label{s}
\end{equation}
where $\Phi_{q+1+2s}$ corresponds to a complete model (Eq. 6) and $\Phi_{q+1}$ corresponds to the algebraic polynomial part. Because the basic functions are not orthogonal, the coefficients $C_\alpha$ are different for both models. The exact coincidence of the observations with the approximation corresponds to $S(f)=1,$ whereas the values at `'bad frequencies'' are typically much smaller.

Even if the preliminary values of the periods of a multi-periodic signal were estimated using one-period approximation, or`'prewhitening'', the final values should be corrected using a complete model (e.g. Andronov and Kudashkina, 1988). The semi-regular variable Z UMa showed two two-harmonic waves, which lead to a complex bias behaviour (Andrych et al., 2020a). The parameters were determined using differential corrections.

Our algorithms are pointed to the period search using trigonometric polynomials of different order with a possible trend, which is approximated by a polynomial of arbitrary order. Such approximations are effective for multi-periodic multi-harmonic signals superimposed on a slow trend. In the software MCV (Andronov and Baklanov, 2004), the approximations may be done for multi-harmonic models for (up to) 3 basic periods with a polynomial trend.

Some stars show fractal-type power-law $S(f)\propto f^{-\gamma}$ (e.g. Andronov et al. 1999, 2008). 

The second type of methods for periodogram analysis is called `'non-parametric''.  Andronov and Chinarova (1997) studied statistical properties of 9 modifications of the test-functions. They were implemented by software by various authors (e.g. Breus, 2007).

The optimal degree of the trigonometrical polynomial $s$ may be determined using the limit for the FAP (False Alarm Probability) (Andronov, 1994a). Kudashkina and Andronov (1996) made an atlas and catalogue of the Fourier characteristics of a group of LPVs (Long Period Variables). Kudashkina and Andronov (2010a) used these characteristics determined for 62 faint Mira-type stars to analyse statistical relations between the phenomenological characteristics of smoothed phase curves. Kudashkina and Andronov (2017) have added `'phase diagrams'', i.e. the dependence of the brightness on its derivative.

Recent reviews on LPVs of different types are presented by Kudashkina (2003, 2019, 2020).

For other class of objects (intermediate polars), Breus et al. (2012, 2013ab, 2019) used a two-period approximation (orbital and spin period) to study the rotational evolution of the magnetic white dwarf. For some stars, only a main wave of the spin period should be taken into account, for others -also a harmonic at a double frequency. For both models, the software MCV was used.

\subsection{Scalegram and Wavelet Analysis}

New effective characteristics of quasi-periodic signals based on the $''\sigma-$ scalegram'' analysis (Andronov, 1997) have been introduced, namely the effective amplitudes, periods (time scales) and slopes of the scalegram.

The main idea is to compute the dependence of the r.m.s. deviation $\sigma(\Delta t)$ of the observations from the fit as a function of the filter half-width $\Delta t.$ 
With an increasing $\Delta t,$ the systematic differences of the approximation from the signal increase, thus one may estimate the effective `'period'' (or cycle length and the amplitude).

The scalegram was applied for additional classification of 173 semi-regular variables (Andronov and Chinarova, 2003).
 
Andronov, Kolesnikov and Shakhovskoy (1997) had found a fractal-type variability in AM Her at time scale from 3 sec to 30 years (7.5 orders of magnitude). Beyond, Andronov (2003) introduced the $''\Lambda-$scalegram analysis, which is some kind of a periodogram analysis.

The wavelet analysis was improved for irregularly spaced data (Andronov, 1998) as a particular case of the scalegram analysis. 
For example, the periodogram and wavelet analysis of the semi-regular variable supergiant Y CVn was presented by Kudashkina and Andronov (2010b) with methodological details.

To increase the accuracy for the studies of period (and other parameters) variations, the ''running sine'' (Andronov and Chinarova, 2013) method was proposed, which is for the signals with high coherence (studied by global approximations) and low coherence (suitable for the wavelet analysis). 
The main idea is to use the `'running approximation''
\begin{equation}
x_C(t;t_0;C_\alpha)=C_1\cdot +C_2\cdot\cos(\omega t)+C_3\cdot\sin(\omega t) 
\end{equation}
only in the `'running'' interval $t_0-\Delta t\le t\le t_0+\Delta t.$ Thus the parameters  $C_\alpha(t_0)$ are functions of $t_0$ and the filter half-width $\Delta t.$ Typically, we choose a `'symmetrical'' value $\Delta t=0.5P,$ whereas, for large observational gaps, it may be enlarged to $\Delta t=1P$ or even more.

This method is effective for either ''nearly-periodic'', or ''modulated periodic'' variations in intermediate polars, pulsating variables etc. 

\subsection{Special Shapes (Patterns)}

For the signals with abrupt changes, a set of approximations using ''special shapes'' was proposed. Particularly, the software NAV (''New Algol Variables'') is effective not only for the EA-type eclipsing variables (Andronov, 2012, Andronov et al., 2012), but also for EB and EW and allows to distinguish these types from non-eclipsing elliptic binaries (Tkachenko et al. 2016) while classifying. Using this phenomenological model for multi-color observations, Andronov et al. (2015) estimated physical parameters of the binary model.

For studies of ''near extremum'' parts of the light curve, including the determination of ToM (Time of Minimum/Maximum), 21 functions (11 types of functions) were realized in the software MAVKA (Andrych and Andronov, 2019,  Andrych et al. 2020ab). Some of the functions were previously introduced by Andronov (2005), Andrych et al. (2015, 2017). These methods were applied to determine ToM of a group of eclipsing variables (e.g. Tvardovskyi et al. 2018, 2020, Tvardovskyi 2019, Kim et al. 2020).

Non-polynomial spline-based functions are used for better approximations of the eclipses (Andronov et al., 2017a) and also for pulsating variables with asymmetric phase curves.

\subsection{Other Methods}

The statistically correct expressions for the auto-correlation functions of detrended signals were presented by Andronov (1994b). They improved previously known expressions for a  removal only of a simple mean (Sutherland et al., 1978).

The Principal Component Analysis (PCA) was discussed by Andronov, Shakhovskoy and Kolesnikov (2003) and Andronov (2003). It was also applied to UBVRI photometry of the asynchronous polar BY Cam (Andronov et al., 2008).

A method for CCD photometry using many stars to improve accuracy was implemented in the program MCV (Andronov and Baklanov, 2004) and its first results were published by Kim et al (2004).

\section{Pulsations of the semi-regular variable RY UMa}

\subsection{Recent data from the AFOEV}

The semi-regular pulsating variable RY UMa (= AN 1909.0001= BD+62 1224 = IRAS 12180+6135
= SAO  015775) is classified in the GCVS as a SRb pulsating variable with a range of brightness variations $6.68^m-8.3^m,$ a period of $P=310^d$ and spectral range M2$-$M3IIIe (Samus et al., 2017).

It was analyzed on 6486 visual (and, partially, CCD) observations from the AFOEV database (http://cdsarc.u-strasbg.fr/afoev/). The number of CCD V observations in this sample is 88 (1.4\%). The time interval HJD 2451629 -- 2458026 continues the previous interval studied in the ''Catalogue of Main Characteristics of Pulsations of 173 Semi-Regular Stars'' (Chinarova and Andronov, 2000), where the periodogram had shown 3 peaks at periods $P=3926^d\pm12^d,$ $303.74^d\pm 0.08^d$ and $285.29^d\pm 0.07^d$ days and corresponding semi-amplitudes $r=0.197^m,$ $0.122^m$ and $0.122^m,$ respectively. 

Our new analysis of the AFOEV database show a single peak with a period $287.00^d\pm 0.14^d,$ much smaller than the GCVS value of $P=310^d$ for the beginning of the XX century, initial epoch for
the maximum brightness (minimum magnitude) $T_0=2454005.2\pm0.8$ and semi-amplitude $r=0.211^m\pm 0.004^m,$ superimposed onto a trend (which was approximated by a parabola). 

These parameters were obtained by using a complete model, without any detrending or prewhitening, as realized in the software MCV (Andronov and Baklanov, 2004). 
The characteristics of the individual maxima and minima were determined using the new version of the software MAVKA (Andrych and Andronov, 2019). 

The brightness at the individual maxima varies from $6.91^m$ to $7.29^m,$ at the minima -- from $7.52^m$ to $8.09^m.$ For the analysis of the smooth variations of the mean brightness (over the cycle of pulsations), semi-amplitude and phase, the ''running sines'' method was applied (Andronov and Chinarova, 2013).

Results are shown in Fig. 1 and are explained in the captions. Except for the `'sine'' approximation, all others show drastic variations in the shape of the individual cycles and the mean brightness. Similar `'switchings'' between the states of `'nearly constant brightness' and oscillations are seen in some other stars, e.g. RU And (Chinarova, 2010).'

\begin{figure}
\includegraphics[width=\textwidth]{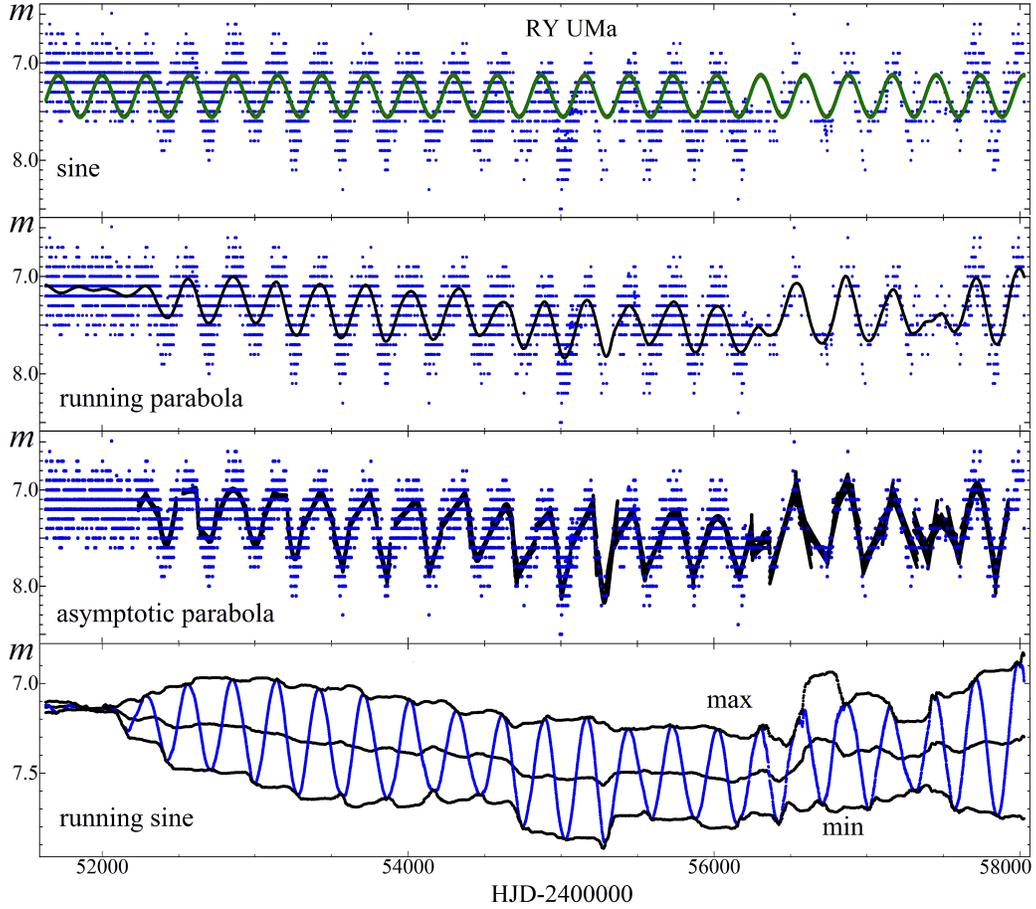}
\caption{Approximations of the light curve of RY UMa from the AFOEV database. The points are the individual data.  The `'sine'' corresponds to the simplest periodic (sine) approximation, i.e. the trigonometrical polynomial of order 1 without any trend (Eq.(6) with $q=0,$ $s=1$). The best fit period is  $P=287.00^d\pm 0.14^d.$  In this model, there are no variations of phase, amplitude or mean brightness (over the pulsation cycle). The `'running parabola'' shows a smooth approximation for all the data, whereas the `'asymptotic parabola'' corresponds to local approximations of separate intervals near extrema.  For the `'running sine'' model, there are shown dependencies of 4 different parameters'': extrapolated maximum (max) and minimum (min) brightness, mean brightness over the pulsation cycle (long-term variations) and the approximation of the current brightness (nearly periodic curve with changing amplitude and other parameters).
}\label{fig:deltapmdec}
\end{figure}

\subsection{Complete data from the AAVSO}

The results from the previous subsection correspond to resent state of the star. As the new period $P=287.00^d\pm 0.14^d.$ significantly differs from $P=310$ mentioned in the GCVS, we have performed an analysis of a complete sample of the observations from the AAVSO database. It contains 38761 observations from JD 2425343 to 2458804, among them 38495 visual and 139 CCD V data. The periodogram $S(f)$ is shown in Fig. 2. It shows 4 peaks at formal periods $9815^d,$ $5147^d,$ $306.4^d$ and $287.5^d.$ The long-term waves seem not to be periodic. The pair of the shorter periods may correspond to a period switch or to simultaneously acting periods causing bias. For these values, the estimates beat period is $P_{\rm beat}\approx 4660^d,$  which is by 10\% different from $5147^d$ seen at the periodogram. 

Figure 3 shows original data, as well as their approximations using different methods. At first, we have used the `'running sine'' approximation reviewed by Andronov and Chinarova  (2013) with the initial light elements mentioned above  $P=287.00^d\pm 0.14^d,$ $T_0=2454005.2\pm0.8.$ The approximation shows changes of the amplitude and the mean level. They are not strictly periodic, but the cycle length is in an agreement with the $5147^d$ value from the periodogram. As was mentioned above, the amplitude of individual pulsations may practically vanish. This occurs, when the star becomes bright.

For an illustration, the variations of the maximum and minimum brightness are shown. Contrary to a sinusoidal shape expected for a beat between two waves with close periods, there is no strict period, but a cycle. 

Next approximation as a `'running parabola''. In Fig. 3, there are shown approximations corresponding to two maxima of the `'signal/noise'' ratio at the filter half-width $\Delta t=4000^d$ and $177^d.$ As expected (Andronov, 1997), these values exceed half of the period/ cycle length.

For long-term variations, one may directly compare the curves marked as `'slow''  for the `'running sines'' and $\Delta t=4000^d$ for the `'running parabolae''. These methods are complementary. For the first method, the amplitude of variations is larger, being more sensitive to statistical fluctuations and gaps in the observations.

In Fig. 4, the variations of the phase of the maximum are shown obtained using the `'running sine'' approximation. This is a scaled version of the typical $(O-C)$ vs $E$ diagram: $(O-C)=\phi\cdot P_0,$ $E=(t-T_0)/P_0.$ The interval of phases $\phi$ was extended from its `'main'' interval $[0,1)$ to wider range to avoid abrupt jumps in phase. 
The phase correction (manual and automatic) by an integer number is available in the software MCV. The nearly straight lines are typical for an abrupt period change. So we have applied the `'asymptotic parabola'' algorithm implemented in MAVKA. The transition time interval between the two lines (different periods) is JD 2447294-2451242, so the duration is $3946^d$ (a dozen per cent accuracy) with a center of the interval at 2449268

After splitting the interval of all data into two subintervals, the following light elements were determined:
\begin{equation}
{\rm Max.JD}=2443914.9(\pm 1.0)+ 307.35(\pm 0.08)\cdot E_1, ~~~~~JD<2449268,
\end{equation}
\begin{equation}
{\rm Max.JD}=2454008.8(\pm 0.5)+ 285.26(\pm 0.06)\cdot E_2, ~~~~~JD\ge 2449268,
\end{equation}
The corresponding semi-amplitudes are $0.162\pm 0.003$ and $0.198\pm0.002.$ They are mean values for these intervals, as this parameter varies in individual pulsational cycles between $0.01^m$ and $0.37^m.$ The r.m.s. deviations of the observations from the approximation $\sigma_0=0.278^m$ and $0.261^m$ are thus much larger than those for the `'running parabola'' approximation $(0.175^m).$
The zero point for the cycle numbering is the closest one to the sample mean of the times of observations, which is obviously different in different intervals.

The ratio of the periods is close to 14/13. Under this assumption, the estimates of the periods are $14\cdot21.95^d=307.3^d$ and $13\cdot21.95^d=285.35^d,$ equal to our results within error estimates. 


An alternative model for the $(O-C)$ variations is a periodic wave superimposed on a linear trend. The best fit corresponds to a long `'period'' of $66pm13$ thousand days $(182\pm36)$yr, which  larger by a factor of two than the duration of observations (91yr), so it can't be confirmed from the present data.

\begin{figure}
\includegraphics[width=\textwidth]{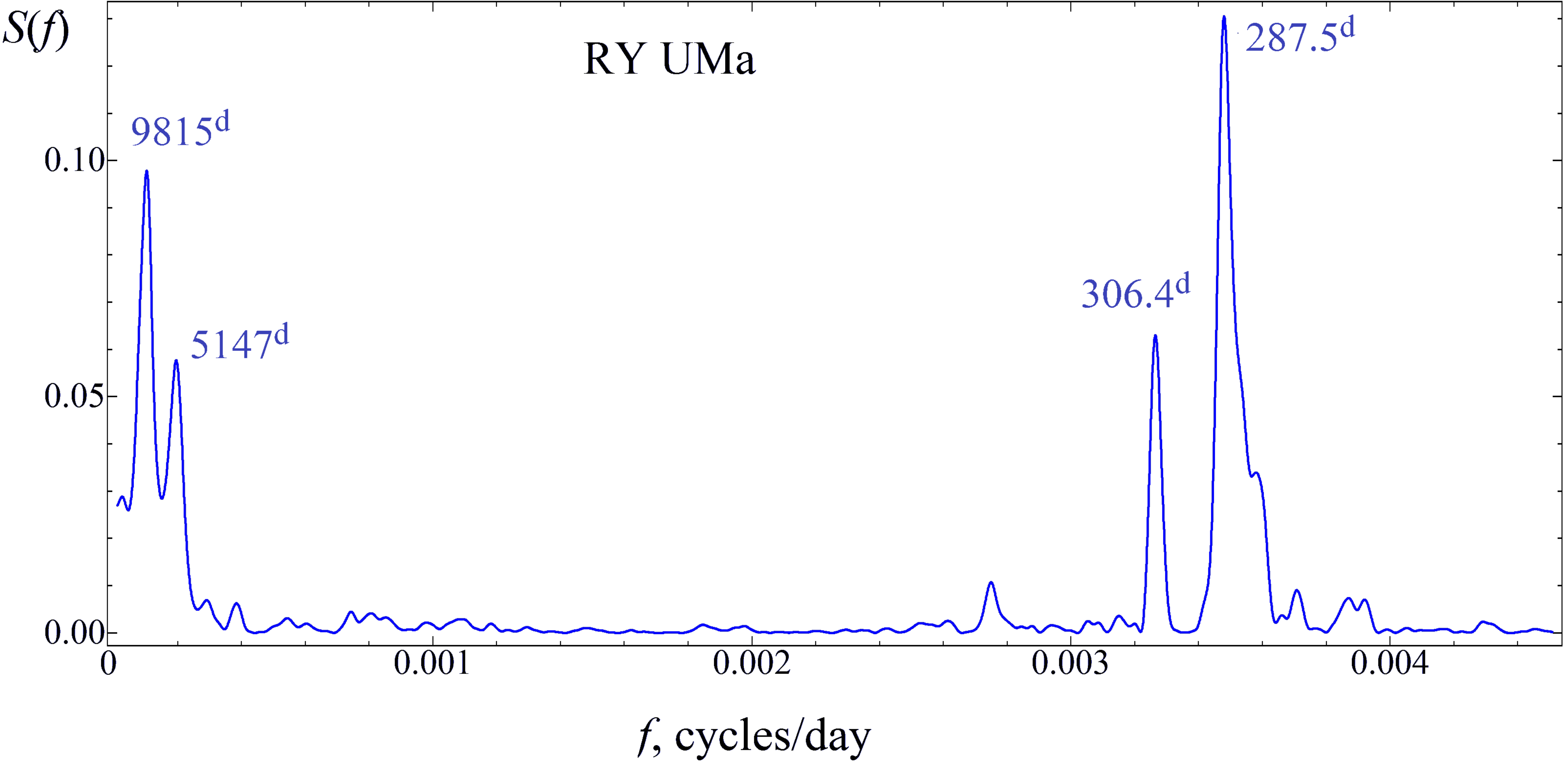}
\caption{The periodogram $S(f)$ of the visual and CCD V observations of  RY UMa from the AAVSO database. The periods corresponding to four highest peaks are shown.
}\label{fig2}
\end{figure}

\begin{figure}
\includegraphics[width=\textwidth]{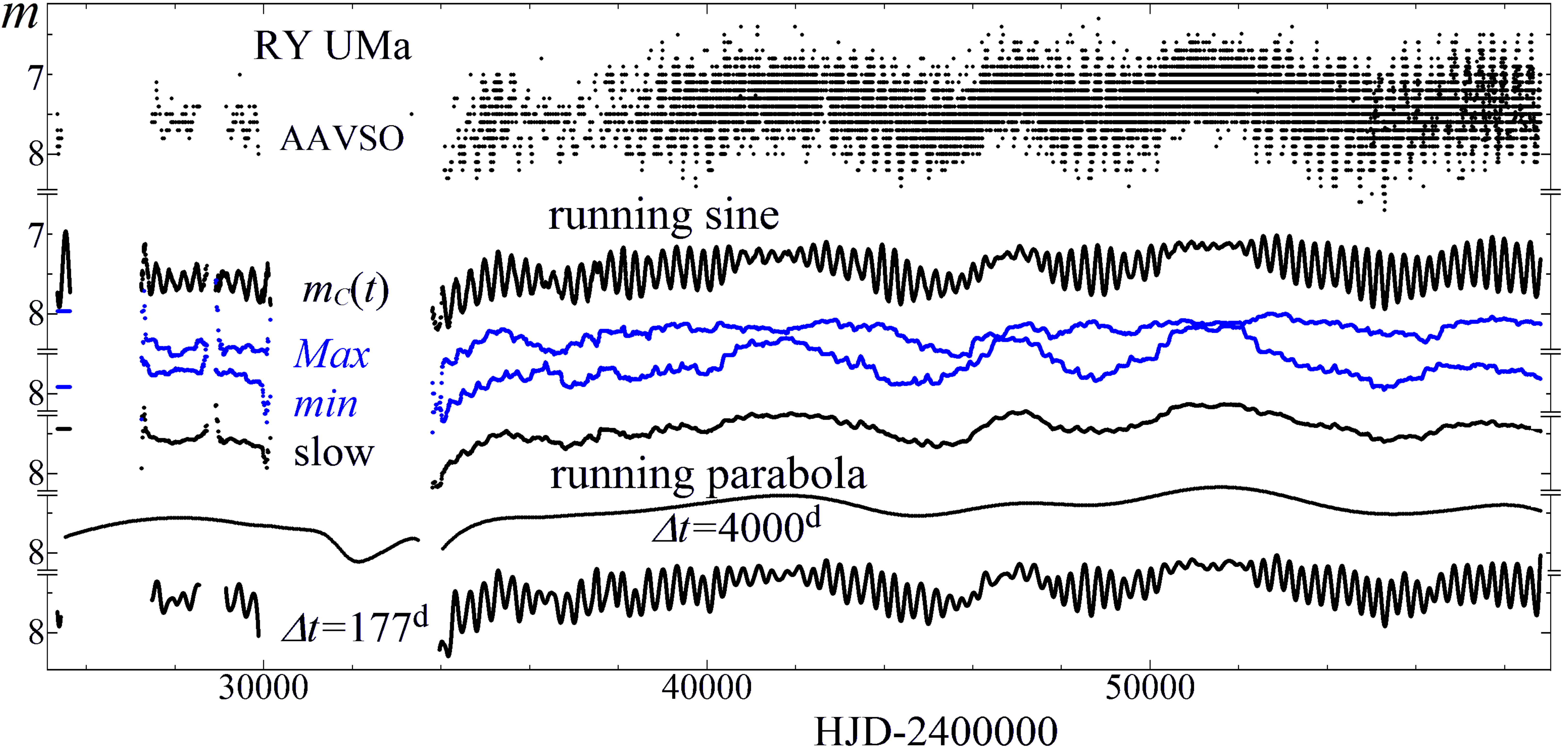}
\caption{Approximations of the light curve of RY UMa from the AAVSO database. The points are the individual data.  The `'sine'' corresponds to the simplest periodic (sine) approximation, i.e. the trigonometrical polynomial of order 1 without any trend (Eq.(6) with $q=0,$ $s=1$). The best fit period is  $P=287.00^d\pm 0.14^d.$  In this model, there are no variations of phase, amplitude or mean brightness (over the pulsation cycle). The `'running parabola'' shows a smooth approximation for all the data, whereas the `'asymptotic parabola'' corresponds to local approximations of separate intervals near extrema.  For the `'running sine'' model, there are shown dependencies of 4 different parameters'': extrapolated maximum (max) and minimum (min) brightness, mean brightness over the pulsation cycle (long-term variations) and the approximation of the current brightness (nearly periodic curve with changing amplitude and other parameters).
}\label{fig3}
\end{figure}

\begin{figure}
\includegraphics[width=\textwidth]{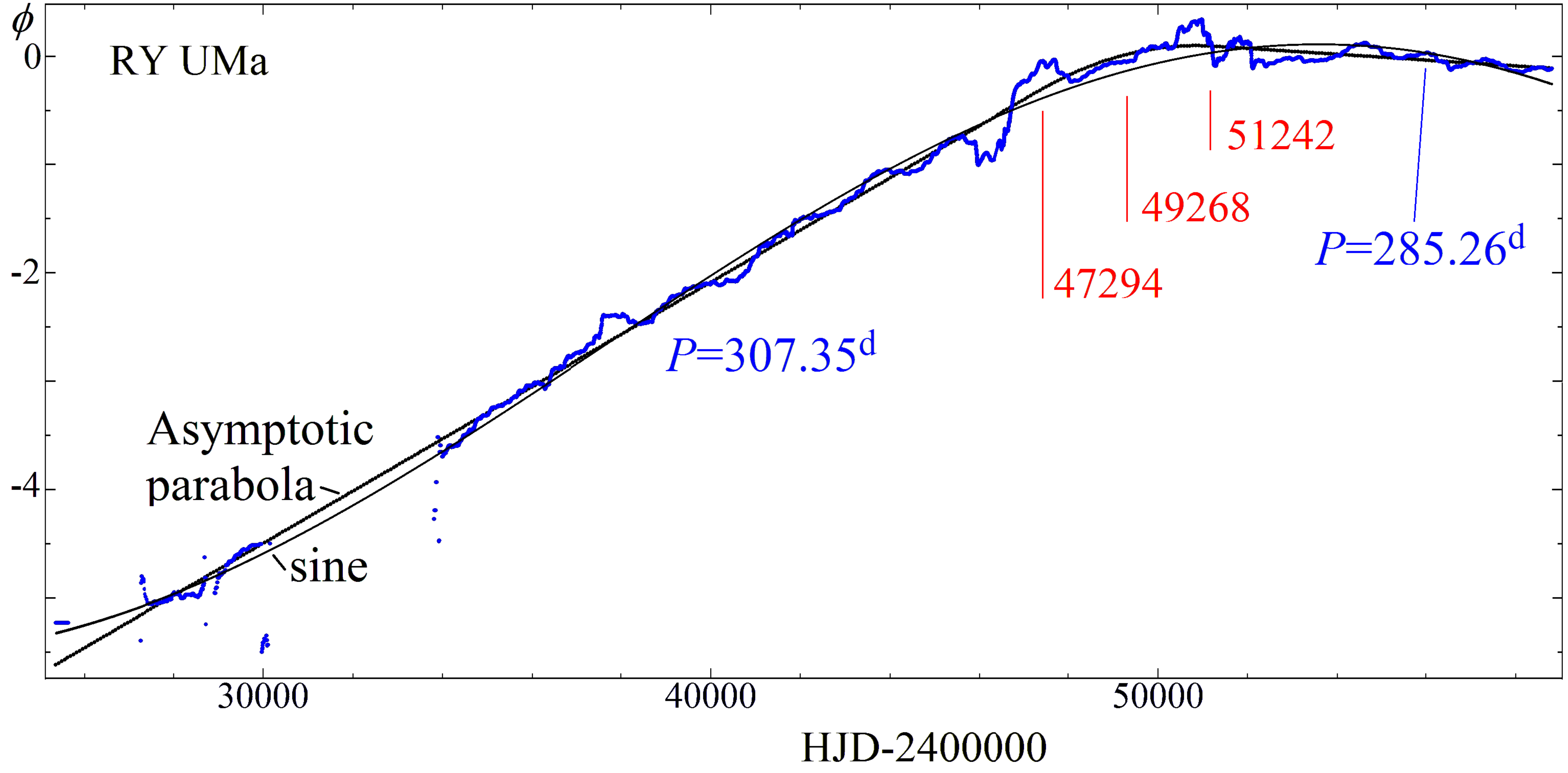}
\caption{Dependence of the phase of maximum of RY UMa (AAVSO) according to the light elements Max.JD~$2454005.2+287^d\cdot E$ computed using the `'running sine''  approximation. The `'sine'' corresponds to the simplest periodic (sine) approximation with a linear trend. The `'asymptotic parabola'' shows a period change. The vertical lines correpond to borders between the intervals of different constant (in a model) period with a link between them. The middle of the internal interval shows the date of the period switch.
}\label{fig4}
\end{figure}

\section{Conclusions}
The methods have been applied (totally) to 2000+ variable stars of different types using our own monitoring, as well as the photometric surveys from ground-based and space observatories. A wide range of types of variability initiated the elaboration of additional methods. They are briefly mentioned with extensive list of links to original papers. 

The methods are illustrated on the light curve of the semi-regular pulsating star RY UMa.
For this star, the period of the dominating wave has changes from $\approx307^d$ before JD 2449268 (in 1993) to $\approx307^d$ after. The characteristic duration of the switch is $\approx4000^d\approx11$yr. The semi-amplitude of the individual cycles drastically varies from $0.01^m$ and $0.37^m,$ which is characteristic for the beat phenomena. The range of variations is $7.0^m-8.2^m$ (running sine approximation). We suggest that pulsations with both periods are present simultaneously, but the amplitudes change with time, thus the apparent period switch indicates the change of the wave, which is dominating in the amplitude.

\section*{ACKNOWLEDGEMENTS}

We thank the French Association of Variable Stars Observers AFOEV (Association Francaise des Observateurs d'Etoiles Variables, http://cdsarc.u-strasbg.fr/afoev) and the American Association of Variable Stars Observers (AAVSO, http://aavso.org) for a huge number of observations of variable stars made by amateurs and available on-line. 
This work is a part of the ''Stellar Bell'' (Andronov et al., 2014) part of the ''Inter-Longitude Astronomy'' (Andronov et al., 2003, 2010, 2017) international project, as well
as of the `'Ukrainian Virtual Observatory'' (Vavilova et al., 2012, Р’Р°РІРёР»РѕРІР° Рё РґСЂ., 2012) and ''AstroInformatics'' (Vavilova et al., 2017).

\subsection*{\rm \bf \normalsize References}
\setlength\parindent{-24pt}
\par

Anderson T.W. (2003) An Introduction to Multivariate Statistical Analysis
3rd ed. - Wiley

Andronov I. L. (1994a) Odessa Astron. Publ. 7, 49

Andronov I. L. (1994b)  Astronomische Nachrichten 315, 353

Andronov I. L.
(1997) A\& A Supplement series, 125, 207 

Andronov I. L. (1998) Kinematika i Fizika Nebesnykh Tel 14, 490

Andronov I. L. (2003) ASP Conf. Ser. 292, 391

Andronov I. L. (2005) ASP Conf. Ser. 335, 37

Andronov I. L. (2012) Astrophysics 55, 536

Andronov I. L. (2020)  Knowledge Discovery in Big Data from Astronomy and Earth Observation: Astrogeoinformatics, ed. P.\v{Skoda} et al., Elsevier (in press)

Andronov I. L. et al. (1999) AJ, 117 (1), 574

Andronov I. L. et al. (2003) AApTr, 22, 793

Andronov I. L. et al. (2008) Central European Journal of Physics 6, 385

Andronov I. L. et al. (2010) Odessa Astronomical Publications, 23, 8

Andronov I. L. et al. (2014) Advances in Astron. Space Phys., 4, 3

Andronov I. L. et al. (2015) JASS, 32, 127

Andronov I. L. et al. (2017) ASP Conf. Ser. 511, 43

Andronov I. L., Baklanov A. V. (2004) Astron. School's Rep., 5, 264

Andronov I. L., Breus V. V., Zola S. (2012) Odessa Astronomical Publications 25, 145

Andronov I. L., Chinarova L.L. (1997)
Kinematics and Physics of Celestial Bodies 13 (N6), 67

Andronov I. L., Chinarova L.L. (2003)
ASP Conf. Ser. 292, 401.

Andronov I. L., Chinarova L. L. (2013)
Czestochowski Kalendarz Astronomiczny, ed. Bogdan Wszolek, 10, 171; arXiv:1308.1129

Andronov I.L., Kolesnikov S.V., Shakhovskoy N.M. 
(1997) Odessa Astronomical Publications 10, 15

Andronov I. L., Kudashkina L. S. (1988)
Astronomische Nachrichten, 309, 323

Andronov I. L., Kudashkina L. S. (2010)
Odessa Astronomical Publications, 23, 67

Andronov I. L., Marsakova V. I. (2006) Astrophysics 49, 370

Andronov I. L., Shakhovskoj N. M., Kolesnikov S. V. (2003)
NATO Science Series II -- Mathematics, Physics and Chemistry, 105, 325

Andrych K. D., Andronov I. L. (2019) Open European Journal on Variable Stars 197, 65

Andrych K. D., Andronov I. L., Chinarova L.L.  (2020) Journal of Physical Studies (subm.)

Andrych K. D., et al. (2015)
Odessa Astronomical Publications, 28, 158

Andrych K. D., et al. (2017)
Odessa Astronomical Publications, 30, 57 

Andrych K.D., et al. (2020a) Journal of Physical Studies, 24, 1902

Andrych K.D., et al. (2020b) Contributions of the Astronomical Observatory Skalnat\'e Pleso, 50, 557

Breus V. V.
(2007) Odessa Astronomical Publications 20, 32

Breus V. V. et al. (2012) Advances in Astronomy and Space Physic, 2, 9-

Breus V. V. et al. (2013a) Journal of Physical Studies 17, 3901

Breus V. V. et al. (2013b) Astrophysics 56, 518

Breus V. V. et al. (2010) Monthly Notices of the Royal Astronomical Society  488, 4526

Chinarova L. L. (2010) Odessa Astron. Publ. 23, 25

Chinarova L. L., Andronov I. L. (2000) Odessa Astron. Publ. 13, 116

Kim Yonggi, et al. (2004) Journal of Astronomy and Space Sciences 21, 191

Kim Yonggi, et al. (2020) Journal of the Korean Astronomical Society, 53, 43

Kudashkina L. S. (2003) Kinematika Fizika Nebesn. Tel, 19, 193

Kudashkina L. S. (2019) Astrophysics 62, 623 

Kudashkina L.S. (2020) AApTr, 31, No. 4 (this volume)

Kudashkina L. S.,  Andronov I. L. (1996) Odessa Astron. Publ. 9, 108

Kudashkina L. S.,  Andronov I. L. (2010a) Odessa Astron. Publ. 23, 65

Kudashkina L. S.,  Andronov I. L. (2010b) Odessa Astron. Publ. 23, 67

Kudashkina L. S.,  Andronov I. L. (2017) Odessa Astron. Publ. 30, 93

Mikulasek Z. (2007) Odessa Astronomical Publications, 20, 138

Samus N.N., Kazarovets E.V., Durlevich O.V., Kireeva N.N., Pastukhova E.N. (2017)
Astron. Rep. 61, 80

Sutherland P. G., Weisskopf M. C., Kahn S. M. (1978)
Astrophysical Journal, 219, 1029.

Tkachenko M. G., et al. (2016) Journal of Physical Studies 20, 4902

Tvardovskyi D. E. (2019) eprint arXiv:1911.12415

Tvardovskyi D. E., et al. (2018) Odessa Astronomical Publications, 30, 103

Tvardovskyi D. E., et al. (2019)  eprint arXiv:1912.02087

Tvardovskyi D. E., et al. (2020)  Journal of Physical Studies, 24, 3902

Tvardovskyi D. E., et al. (2019) eprint arXiv:1912.02087

Vavilova I. B. et al. (2012)  Kinematics and Physics of Celestial Bodies, 28, 85

Vavilova I. B. et al. (2017) Proc. IAU Symposium, 325, 361

Вавилова И.Б. et al. (2012)  Кинематика и физика небесных тел, 28, 59

Вавилова И.Б., Пакуляк Л.К. Пакуляк, Процюк Ю.И., и др. (2011)
Космическая наука и технология, 17, 74

Кудашкина Л.С. (2003) Кинематика и физика небесных тел 19 (3), 193

\end{document}